\pdfoutput=1
\documentclass[aps,preprint,groupedaddress,nofootinbib]{revtex4-1}
\usepackage[]{natbib}
\usepackage{amsmath, amssymb, amsthm, url}
\usepackage{graphicx}
\usepackage[table]{xcolor}  
\usepackage[normalem]{ulem} 
\usepackage{dcolumn}
\usepackage{setspace}
\usepackage{booktabs}
\newcolumntype{d}[1]{D{.}{.}{#1}}

\usepackage{booktabs}

\newcolumntype{x}[1]{>{\centering\arraybackslash\hspace{0pt}}m{#1}}
\newcolumntype{.}{D{.}{.}{3.0}}

\begin{document}


\title{Temperate and chronic virus competition leads to low lysogen frequency}
\spacing{1}


\author{Sara M.~Clifton}
\email{clifto2@stolaf.edu}
\affiliation{Department of Mathematics, Statistics, and Computer Science, St.~Olaf College, Northfield, Minnesota 55057, USA}

\author{Rachel J.~Whitaker}
\affiliation{Department of Microbiology, University of Illinois at Urbana-Champaign, Urbana, Illinois 61801, USA}
\affiliation{ Carl R.~Woese Institute for Genomic Biology, University of Illinois at Urbana-Champaign, Urbana, Illinois 61801, USA}

\author{Zoi Rapti}
\affiliation{Department of Mathematics, University of Illinois at Urbana-Champaign, Urbana, Illinois 61801, USA}
\affiliation{ Carl R.~Woese Institute for Genomic Biology, University of Illinois at Urbana-Champaign, Urbana, Illinois 61801, USA}



\begin{abstract} 
The canonical bacteriophage is obligately lytic: the virus infects a bacterium and hijacks cell functions to produce large numbers of new viruses which burst from the cell. These viruses are well-studied, but there exist a wide range of coexisting virus lifestyles that are less understood. Temperate viruses exhibit both a lytic cycle and a latent (lysogenic) cycle, in which viral genomes are integrated into the bacterial host. Meanwhile, chronic (persistent) viruses use cell functions to produce more viruses without killing the cell; chronic viruses may also exhibit a latent stage in addition to the productive stage. Here, we study the ecology of these competing viral strategies. We demonstrate the conditions under which each strategy is dominant, which aids in control of human bacterial infections using viruses. We find that low lysogen frequencies provide competitive advantages for both virus types; however, chronic viruses maximize steady state density by eliminating lysogeny entirely, while temperate viruses exhibit a non-zero `sweet spot' lysogen frequency. Viral steady state density maximization leads to coexistence of temperate and chronic viruses, explaining the presence of multiple viral strategies in natural environments.
\end{abstract}

\maketitle



\section{Introduction}
All viruses depend upon their hosts for reproduction. Viruses have evolved many strategies to reproduce within bacteria, including the lytic, temperate, and chronic lifestyles within the bacterial host \cite{calendar2006bacteriophages,dimmock2016introduction,weinbauer2004ecology}. After infection, lytic viruses replicate within the bacterial host and transmit by bursting from the cell, killing the host. Temperate viruses have both a lytic cycle and a latent cycle, in which the viral genetic material is integrated into host genomes; latent viruses remain dormant in the bacterial genome until induced to replicate \cite{weinbauer2004ecology}. In chronic infection, productive host cells bud new viruses from the cell without killing the bacterium \cite{rakonjac2012}. Chronic viruses may also have a latent cycle in which viral genetic material is incorporated into the bacterium's genome, and the cell transmits the virus's genetic material (provirus) to daughter cells vertically \cite{lwoff1953lysogeny}. Comparative genomics among closely related bacterial strains has uncovered a plethora of proviruses of both temperate and chronic lifestyles \cite{davies2016role,roux2015viral,mathee2008dynamics,mosquera2016pangenome,spencer2003whole,kung2010accessory}. 

Viruses of all four lifestyle classes infect many bacteria relevant to human disease treatment, especially immunocompromised patients vulnerable to common bacterial infections. In particular, patients with cystic fibrosis or serious burns may become infected with the ubiquitous \textit{Pseudomonas aeruginosa}, which can lead to patient death within days if unsuccessfully treated \cite{courtney2007predictors,emerson2002pseudomonas,nixon2001clinical}. 
Because \textit{P.~aeruginosa} is often resistant to multiple antibiotic treatments \cite{jarvis1992predominant,hancock2000antibiotic,us2013antibiotic}, phage therapy \cite{altamirano2019phage,sulakvelidze2001bacteriophage} and phage-antibiotic synergistic (PAS) therapy \cite{lin2018synergy,comeau2007phage,kutter2010phage} are now being studied to treat bacterial infections. 
Response to these treatments depends significantly on the ecology of the bacteria-virus system already present within the human host \cite{clifton2019modeling}; therefore it is critical to understand the environmental and evolutionary conditions under which each viral strategy is dominant in order to provide effective treatment.

While mathematical models of lytic viruses (e.g., \cite{weitz2008,payne2001}) and temperate viruses (e.g.,  \cite{sinha2017silico}) have been studied extensively, relatively few models of chronic viruses (e.g., \cite{gulbudak2019heterogeneous,weitz2019viral,clifton2019modeling}) have been examined. To our knowledge, no studies have rigorously analyzed the ecological interactions among bacteria and all four viral lifestyles: lytic, latent lytic, chronic, and latent chronic. Several important open questions exist that neither experimental nor modeling efforts have yet answered in this context:
\begin{enumerate}
	\item Experiments have found that temperate virus lysogen frequencies tend to be small ($\sim$1\% of infections) \cite{calendar2006bacteriophages}. What are the theoretical underpinnings of this phenomenon?
	\item Lysogen frequencies for temperate viruses have been well-studied both experimentally and theoretically \cite{calendar2006bacteriophages,oppenheim2007new,volkova2014modeling}, but lysogen frequencies for chronic viruses have not been determined. What is the predicted range of lysogen frequencies for chronic viruses?
	\item Bacterial recovery (cure) rates from viral infection have not been quantified either experimentally or theoretically. Most mathematical models ignore recovery for simplicity (e.g., \cite{weitz2019viral}). Experimentally, a wide range of recovery rates has been observed; some proviruses remain viable over evolutionary timescales (implying recovery rates near zero) \cite{brussow2004phages}, and some proviruses are inactivated nearly instantly by CRISPR systems (implying extremely fast recovery rates) \cite{horvath2010crispr}. Can we establish a narrower range of typical recovery rates?
\end{enumerate}

In this paper we develop a mathematical model of the competition between temperate and chronic viruses for bacterial hosts, using \textit{P.~aeruginosa} infections within humans as motivation. An analysis of this model yields simple and intuitive answers to the preceding questions.

\section{Model}
\begin{figure}
  \centering
    \includegraphics[width=\textwidth]{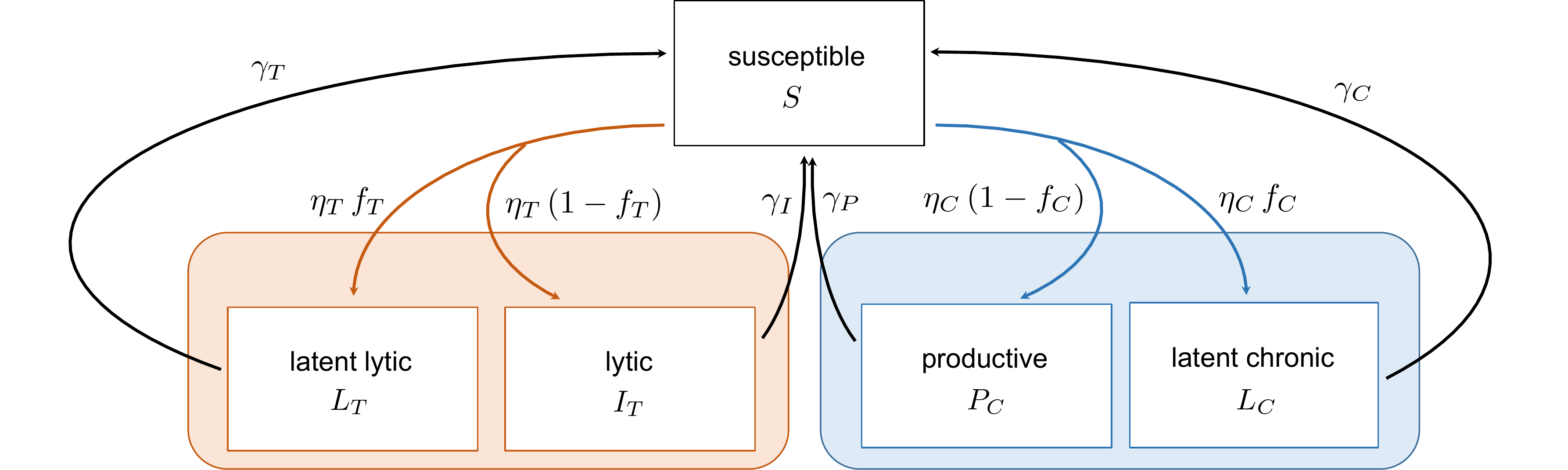}
      \caption{Flowchart of model equations (\ref{eq:sus}-\ref{eq:chronvir}). Arrows indicate infection by temperate viruses (orange), infection by chronic viruses (blue), or recovery from infection (black). Infection rates are denoted $\eta_i$, recovery (cure) rates are denoted $\gamma_i$, and lysogen frequencies are denoted $f_i$.} \label{fig:flowchart}
\end{figure}

With the goal of understanding competition between two viral strategies, temperate viruses $V_T$ with lytic and latent lytic stages and chronic viruses $V_C$ with productive and latent chronic stages, we develop a simple model of the bacteria-virus ecosystem (see Figure \ref{fig:flowchart} for the model overview). Each virus may infect a single strain of bacteria that is initially susceptible ($S$) to both viral types. We assume the total bacterial population $N$ grows logistically to a carrying capacity $K$ \cite{zwietering1990modeling}. Susceptible bacteria grow at a rate $r_S$, and infected bacteria may grow at either faster or slower rates \cite{shapiro2016evolution}.

Temperate viruses infect susceptible bacteria at a rate $\eta_T$; the infected bacteria will either become latently infected $L_T$ with probability $f_T$, or will enter a lytic state $I_T$ with probability $1-f_T$. Bacteria in the lytic state produce viruses and burst (with burst size $\beta_T)$ at a rate $\delta$. While in the lytic state, the virus hijacks cell functions, and the cell cannot reproduce \cite{tabib2017full,st2008determination}. Bacteria in the latent lytic state reproduce at a rate $r_T$. Bacteria do not move between lytic and latent states unless the system is stressed (e.g., by heat or sublethal antibiotics); we ignore spontaneous induction because it is a rare occurrence \cite{nanda2015impact,garro1974relationship,Cortes546275}. However, lytic and latent lytic bacteria may recover from infection at rate $\gamma_I$ and $\gamma_T$, respectively \cite{casjens2003prophages,brussow2004phages}.

Similarly, chronic viruses infect susceptible bacteria at a rate $\eta_C$, leading to either latent infection $L_C$ with probability $f_C$ or productive infection $P_C$ with probability $1-f_C$ \cite{hobbs2016diversity}. Bacteria in the productive state reproduce at a rate $r_P$ and produce viruses at a rate $\beta_C$ without cell death. While in the latent chronic state, bacteria reproduce at a rate $r_C$. Unless the system is stressed, bacteria will not switch from latent to productive states, but productive and latent chronic bacteria may recover at rate $\gamma_P$ and $\gamma_C$, respectively \cite{casjens2003prophages,brussow2004phages}. Again, we ignore spontaneous induction due to its rarity \cite{nanda2015impact,garro1974relationship,Cortes546275}.

Once a bacterium is infected, we assume it will exclude both superinfection by the same viral type and cross infection by viruses of the other type \cite{de2017pseudomonas}. Outside the cell, free temperate viruses and free chronic viruses decay naturally at rates $\mu_T$ and $\mu_C$, respectively \cite{heldal1991production}. 

The following equations (\ref{eq:sus}-\ref{eq:chronvir}) are the dynamical systems model that captures the preceding qualitative description. See Table \ref{tab:variable} for variable descriptions and Table \ref{tab:param} for parameter descriptions and relevant values.

\begin{align}
& \dot{S} = \underbrace{r_S S \left( 1-\frac{N}{K} \right)}_{\text{growth}} - \underbrace{\eta_T S V_T - \eta_C S V_C}_{\text{infection}} + \underbrace{\gamma_T L_T + \gamma_I I_T+ \gamma_P P_C + \gamma_C L_C}_{\text{recovery}} 
\label{eq:sus} \\
& \dot{I}_T = \underbrace{\eta_T (1-f_T) S V_T}_{\text{infection}} - \underbrace{\delta I_T}_{\text{lysis}} - \underbrace{\gamma_I I_T}_{\text{recovery}}
\label{eq:lyt} \\
& \dot{L}_T =   \underbrace{r_T L_T \left( 1-\frac{N}{K} \right)}_{\text{growth}} + \underbrace{\eta_T f_T S V_T}_{\text{infection}} - \underbrace{\gamma_T L_T}_{\text{recovery}}
\label{eq:lytlat} \\
& \dot{P}_C = \underbrace{r_P P_C \left( 1-\frac{N}{K} \right)}_{\text{growth}} + \underbrace{(1-f_C) \eta_C S V_C}_{\text{infection}} - \underbrace{\gamma_P P_C}_{\text{recovery}}
\label{eq:chron} \\
& \dot{L}_C =  \underbrace{r_C L_C \left( 1-\frac{N}{K} \right)}_{\text{growth}} + \underbrace{f_C \eta_C S V_C}_{\text{infection}} -\underbrace{\gamma_C L_C}_{\text{recovery}}
\label{eq:chronlat}\\
& \dot{V}_T = \underbrace{\beta_T \delta I_T}_{\text{burst}} - \underbrace{\eta_T S V_T}_{\text{adsorption}} - \underbrace{\mu_T V_T}_{\text{degradation}}
\label{eq:lytvir} \\
& \dot{V}_C = \underbrace{\beta_C P_C}_{\text{production}} - \underbrace{\eta_C S V_C}_{\text{adsorption}} - \underbrace{\mu_C V_C}_{\text{degradation}}
\label{eq:chronvir}
\end{align}

	\begin{table}[!ht]
	\caption{Description of model variables in bacteria-virus system (\ref{eq:sus}-\ref{eq:chronvir}). Due to nondimensionalization of density and time, all variables and parameters are nondimensional; all densities are relative to the bacterial carrying capacity and all rates are relative to the growth rate of uninfected bacteria.}
\footnotesize
	\begin{tabular}{| c  p{7.5cm} |}  \hline 
	{\bf Variable} & {\bf Meaning} \\  \hline 
	$S$                  & density of susceptible bacteria  \\
	$I_T$               & density of lytic bacteria preparing to burst \\
	$L_T$              & density of latent lytic bacteria     \\
	$P_C$             & density of productive bacteria \\
	$L_C$              & density of latent chronic bacteria \\
	$N$                  & density of all bacteria ($S + I_T +L_T +P_C + L_C$)  \\
	$V_T$             & density of free temperate viruses \\
	$V_C$             & density of free chronic viruses  \\
	$t$                    & time normalized by bacterial reproduction rate \\
	\hline
	\end{tabular}
	\label{tab:variable}
	\end{table}

	\begin{table}[!ht]
	\caption{Description of model parameters in bacteria-virus system (\ref{eq:sus}-\ref{eq:chronvir}). Due to nondimensionalization of density and time, all variables and parameters are nondimensional; all densities are relative to the bacterial carrying capacity and all rates are relative to the growth rate of uninfected bacteria.}
	\begin{center} 
\footnotesize
	\begin{tabular}{| c  p{7.8cm} p{1.8cm}  p{1.4cm} p{1.4cm} |}  \hline 
	{\bf Parameter} & {\bf Meaning} & {\bf Range$^a$} & {\bf Baseline} & {\bf Sources} \\  \hline 
	$r_S$ & net growth rate of susceptible bacteria, normalized to 1 &  1$^b$ & 1 & \cite{spalding2018mathematical} \\ 
	$r_T, r_P, r_C$ & growth rates of (respectively) latent lytic, productive, and latent chronic bacteria & [0.5, 3]$^c$ & 1 & \cite{shapiro2016evolution,kopf2016trace} \\ 
	$K$ & carrying capacity of bacteria, normalized to 1 & 1 & 1 & \\  
	$\eta_T, \eta_C$ & infection rate of (respectively) temperate and chronic viruses  & [0.38, 14.7]$^d$  & 1  & \cite{sinha2017silico} \\ 
	$\gamma_T, \gamma_P, \gamma_C$ & recovery rates of (respectively) latent lytic, productive, and latent chronic bacteria  & [0,1000]$^e$  & 0.67$^f$  & \cite{brussow2004phages,horvath2010crispr} \\
	$\gamma_I$ & recovery rates of lytic bacteria  & [0,1000]  & 0$^g$  & \cite{brussow2004phages,horvath2010crispr} \\ 
	$\delta$ & rate at which lytic infection leads to bursting (eclipse and rise phase) &  [1.5, 7.8]$^h$ & 4 & \cite{yu2017characterization,el2015isolation} \\ 
	$f_T$ & lysogen frequency for temperate viruses & [0, 0.9]  & 0.01  & \cite{calendar2006bacteriophages,oppenheim2007new,volkova2014modeling} \\ 
	$f_C$ & lysogen frequency for chronic viruses & [0, 0.9]  & 0$^f$  & \cite{calendar2006bacteriophages,oppenheim2007new,volkova2014modeling} \\ 
	$\beta_T$ & burst size for bacteria infected with $V_T$ &  [10, 1000] & 100 & \cite{yu2017characterization,el2015isolation,latino2014novel,schrader1997bacteriophage,ceyssens2010molecular,garbe2011sequencing,you2002effects}  \\ 
	$\beta_C$ & viral production rate for bacteria infected with $V_C$ & [5, 200] & 20 & \cite{clifton2019modeling}   \\ 
	$\mu_T, \mu_C$ & degradation rate of (respectively) free temperate viruses and free chronic viruses &  [0.9, 3.6]$^i$ & 1  & \cite{heldal1991production}  \\ 
	\hline
	\end{tabular}
	 \end{center}
	 \begin{flushleft}
	$^a$all parameter ranges are taken for the human pathogens \textit{P.~aeruginosa} or \textit{E.~coli} and their viruses, unless otherwise noted. \\
	 $^b$growth rate is approximately 5.1e-3 min$^{-1}$ for \textit{P.~aeruginosa} grown \textit{in vitro}. \\
	 $^c$estimates based on \textit{E.~coli} and M13 phage.  \\
	 $^d$estimates based on \textit{E.~coli} and $\lambda$ phage.  \\
	 $^e$a wide range of recovery rates has been found; some proviruses are viable over evolutionary timescales and some proviruses are inactivated nearly instantly by CRISPR systems. \\
	 $^f$estimated from viral steady state density (see Results section). \\
	  $^g$selected to be 0 to simplify model analysis; allowing $\gamma_I = \gamma_T$ produces qualitatively similar results, so the increased model complexity is not justified.\\
	  $^h$low estimate is for PAXYB1 phage and PAO1 host, high estimate is for PAK\_P3 phage and PAO1 host. \\
	  $^i$low estimate is for viruses extracted from Raunefjorden, high estimate is for viruses extracted from Bergen Harbor (strains unknown).
	  \end{flushleft}
	\label{tab:param}
	\end{table}

Note that infection rates for temperate and chronic viruses, although assumed constant, may depend on the bacterial population; many relevant bacteria form biofilms at high density that protect the population from infection \cite{harper2014bacteriophages}. For simplicity, we have also assumed that lysogen frequencies are constant, but some studies have demonstrated that bacterial density may impact lysogeny rates \cite{hargreaves2014does,silpe2018host}. These simplifications are necessary for analytic tractability.

\section{Results}
Although this model is applicable to any bacteria-virus ecosystem with both temperate and chronic viral lifestyles, we present model predictions using parameters taken from the human pathogens \textit{P.~aeruginosa} or \textit{E.~coli} and their viruses (see Table \ref{tab:param}). Besides their medical relevance, \textit{P.~aeruginosa} and \textit{E.~coli} are among the most well-studied microbes in science. We initialize the system with $S(0)=1\mathrm{e}{-3}, V_T(0)=V_C(0)=1\mathrm{e}{-7}$ and all others zero, following \cite{sinha2017silico}; no bistability exists for our parameter values, so the initial condition does not affect the steady state (see Supplementary Information). Because we are primarily interested in recovery from infection that is passed vertically to daughter cells, we take $\gamma_I= 0$ for the remainder of the paper. The results are qualitatively similar for $\gamma_I > 0$; see Supplementary Information for full model analysis.

\subsection{Model behavior}
\begin{figure}
  \centering
    \includegraphics[width=\textwidth]{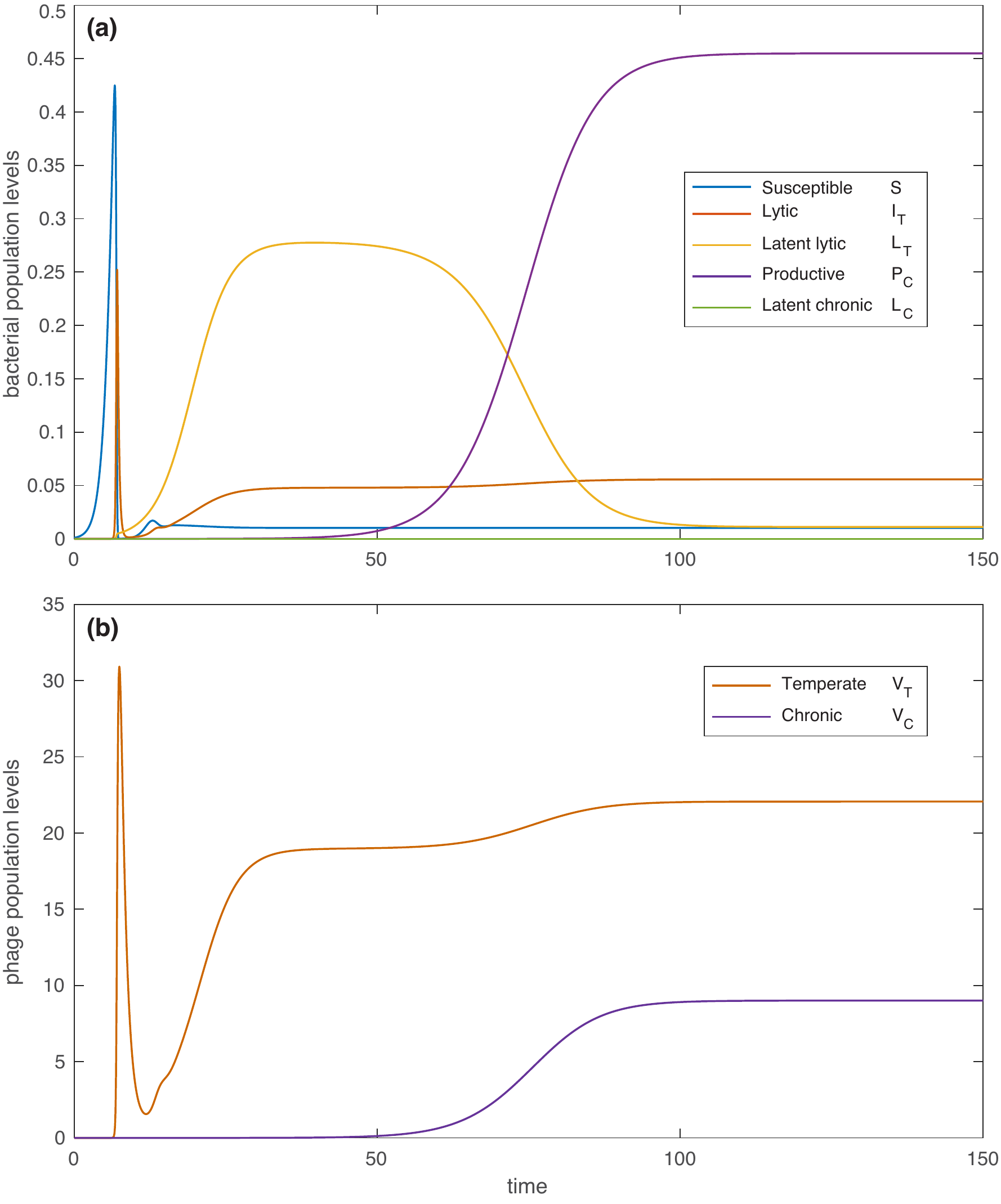}
      \caption{Simulation of model equations (\ref{eq:sus}-\ref{eq:chronvir}). Initial conditions are $S(0)=1\mathrm{e}{-3}, V_T(0)=V_C(0)=1\mathrm{e}{-7}$ with all others zero. All parameters are held constant at the baseline values given in Table \ref{tab:param}. Eventually the temperate and chronic viral strategies reach a stable coexistence state.} \label{fig:sim1}
\end{figure}

In Figure \ref{fig:sim1}, we simulate the model system for the baseline parameter values in Table \ref{tab:param}. We see a rapid initial growth of the susceptible population followed quickly by a population crash caused primarily by lytic infection. Filling the niche created by the susceptible population crash are the latent lytic bacteria. After dozens of bacterial divisions, the latent lytic bacteria recover from infection create a niche for chronically infected bacteria. Eventually, the chronically infected bacteria become the most abundant in the system because chronic viruses do not require new susceptible bacteria in order to reproduce. 

Although chronically infected bacteria dominate the system, free temperate viruses stabilize at over twice the density of free chronic viruses (Figure \ref{fig:sim1}b). Although little is known about the proportion of each viral type seen in natural environments, it is known that temperate and chronic viruses frequently coexist \cite{winstanley2009newly}. The model predicts that the total virus to total bacteria ratio stabilizes at 58:1, which falls within the typical range of virus to bacteria ratios seen in natural environments \cite{james2015,knowles2016lytic}. 

The model behaviors presented here use the baselines in Table \ref{tab:param}. However, the equilibria and their respective stability depends on nearly all the model parameters (see the Supplementary Information for more details).

\subsection{Steady states and stability}
While many evolvable parameters (such as viral burst sizes and bacterial growth rates) are likely limited by physical constraints, lysogen frequency could theoretically take on any value. Lysogen frequency is of particular interest because latency involves inherent tradeoffs between vertical and horizontal transmission; the lytic strategy relies on horizontal transmission only, while the latent strategy uses only vertical transmission.  

\begin{figure}[h]
  \centering
    \includegraphics[width=0.8\textwidth]{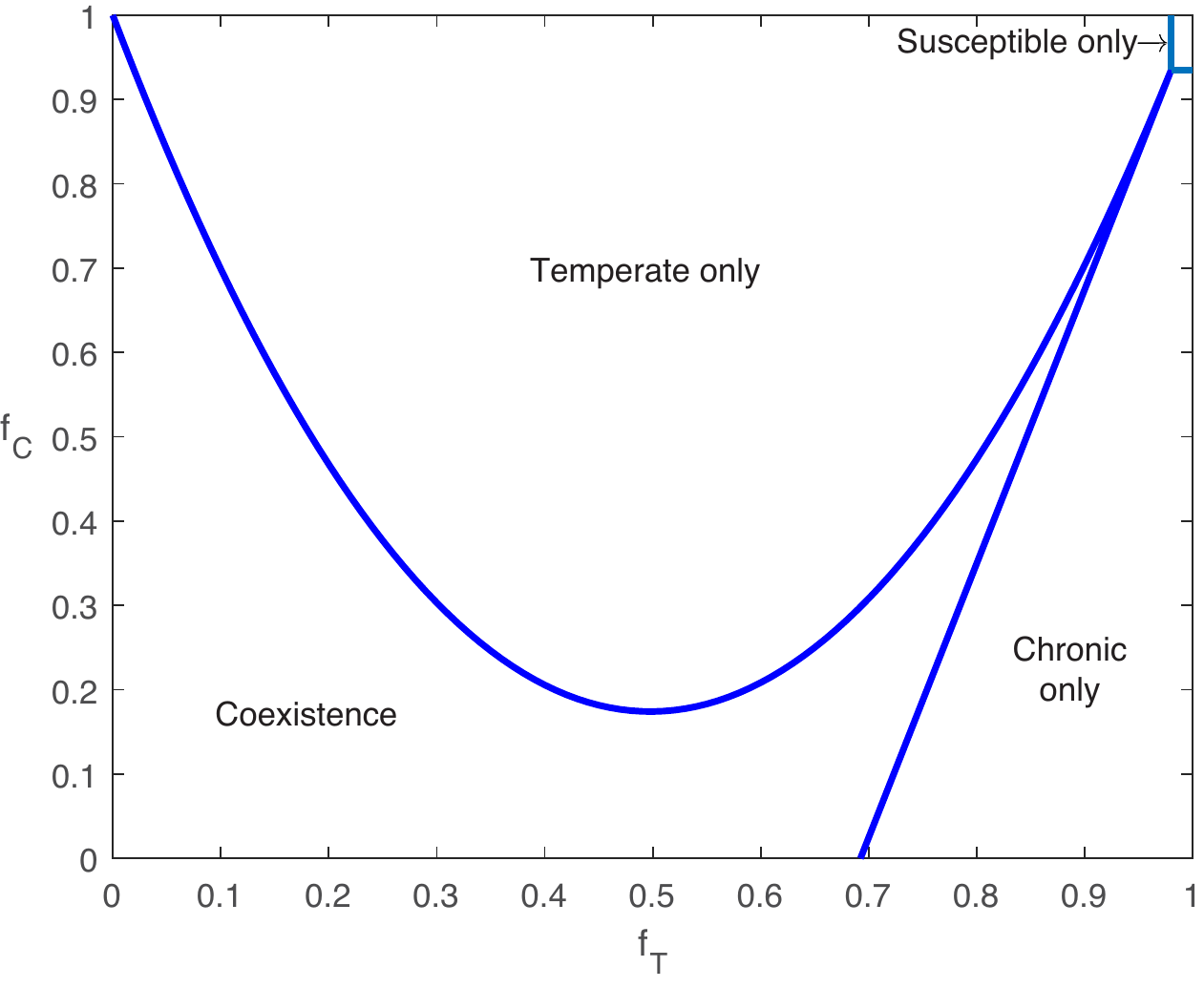}
      \caption{Bifurcation diagram of lysogen frequencies for temperate and chronic viruses. Holding all other parameters constant at the baseline values given in Table \ref{tab:param}, we find that four steady state outcomes are possible: coexistence of both viral strategies, temperate survival with chronic extinction, chronic survival with temperate extinction, and extinction of both viral strategies (susceptible bacteria the only survivors). If the lysogen frequency for a particular viral type is too high, then that virus will not survive. If lysogen frequencies are low enough for both viral types, then the viruses will coexist. This bifurcation diagram was generated using standard linear stability analysis; see the Supplementary Information for details on the viral invasion analysis.} \label{fig:bif3}
\end{figure}

Therefore, our primary interest for this study is the lysogen frequencies for temperate and chronic viruses; we look at the possible steady state outcomes for all possible combinations of lysogen frequencies $f_T$ and $f_C$ with all other parameters held constant at the baselines in Table \ref{tab:param}. Only four stable steady states exist: \textbf{coexistence} (all populations exceed 0), \textbf{temperate strategy only} ($V_C=P_C=L_C=0$), \textbf{chronic strategy only} ($V_T=I_T=L_T=0$), and \textbf{susceptible only} (all populations, except $S$, are 0). See Figure \ref{fig:bif3} for the bifurcation diagram.

Temperate viruses can outcompete chronic viruses if the temperate lysogen frequency is neither too large nor too small (see `Temperate only' region of Figure \ref{fig:bif3}). If the temperate lysogen frequency is too high, then chronic viruses will be produced in large numbers relative to temperate; chronic viruses will infect susceptible bacteria first, eventually driving temperate viruses to extinction (see `Chronic only' region of Figure \ref{fig:bif3}). If the temperate lysogen frequency is too low, then temperate viruses lyse bacteria too quickly, leaving room for productive bacteria to reproduce while the susceptible population also grows; if the productive bacterial population is large enough (i.e., chronic lysogen frequencies are sufficiently low), then both viral types will coexist (see `Coexistence' region of Figure \ref{fig:bif3}). See the Supplementary Information for the complete steady state analysis.

\subsection{Temperate viruses exhibit a `sweet spot' lysogen frequency}
Temperate viruses are known to exhibit a small, nonzero lysogen frequency \cite{calendar2006bacteriophages,oppenheim2007new,volkova2014modeling,yu2017characterization,el2015isolation,latino2014novel,schrader1997bacteriophage,ceyssens2010molecular,garbe2011sequencing,you2002effects}. Our model illustrates why temperate viruses have a theoretical `sweet spot' lysogen frequency when competing with chronic viruses. 

Suppose that temperate viruses select\footnote{We use `select' in the sense of evolutionary game theory.} a lysogen frequency $f_T$ that maximizes steady state viral abundance, including both free viruses ($V_T$) and proviruses ($L_T$, as proxy). Because chronic viruses may also select a lysogen frequency $f_C$ that maximizes their viral density, the optimal lysogen frequency for temperate viruses depends on $f_C$. 

\begin{figure}[h]
  \centering
    \includegraphics[width=\textwidth]{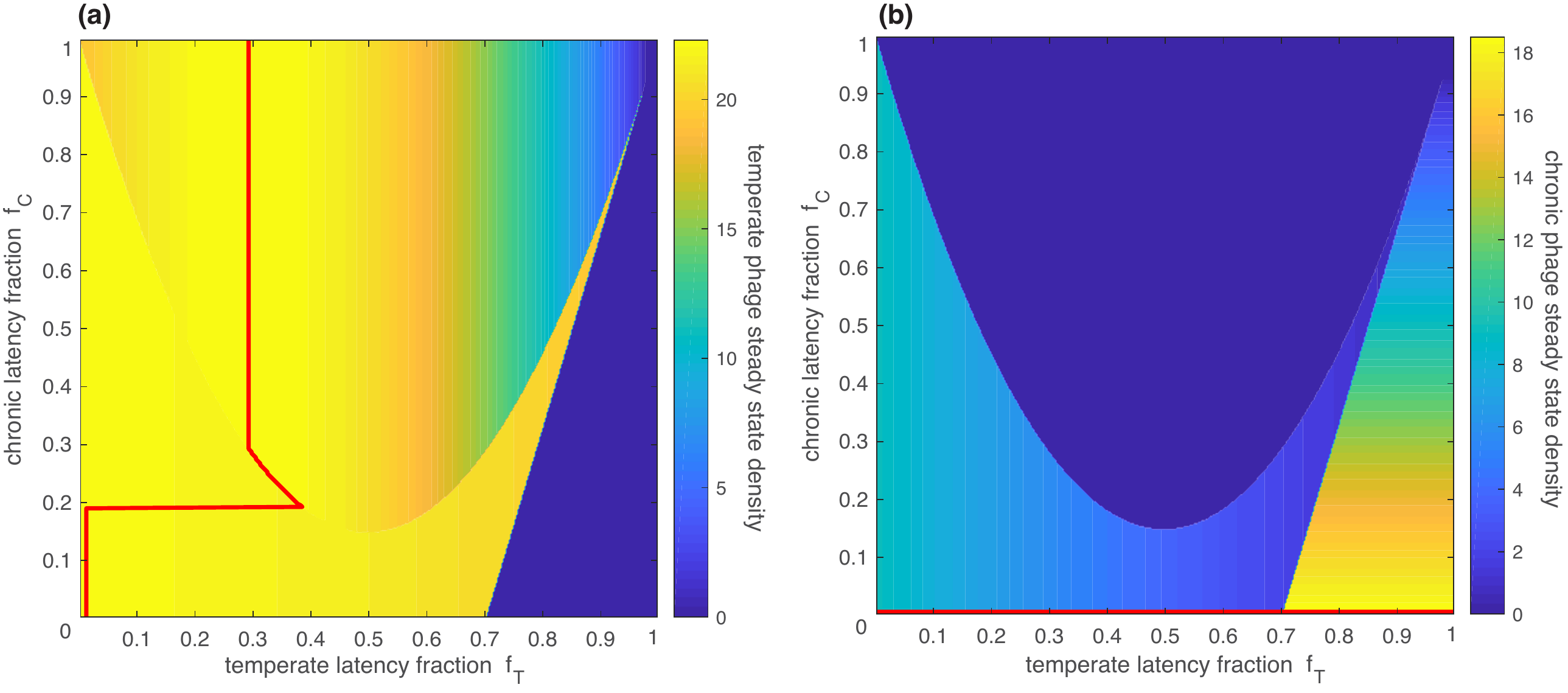}
      \caption{Viral steady state density over the full range of possible lysogen frequencies ($0\le f_i \le 1$) for temperate and chronic viruses. Color indicates the steady state viral density, and the red line is the maximum steady state density for each lysogen frequency of the competing virus. \textbf{(a)} Temperate virus density ($V_T+L_T$) at steady state. The red curve shows the maximum steady state density for every possible chronic lysogen frequency. Note that the optimal temperate lysogen frequency jumps from nearly 40\% lysogeny to about 1\% lysogeny when the chronic lysogen frequency drops below 20\%. This rapid transition occurs because the optimal state for the temperate virus jumps from a temperate only state to a coexistence steady state. \textbf{(b)} Chronic virus density ($V_C+L_C$) at steady state. The red curve shows that the maximum steady state density for every possible chronic lysogen frequency is $f_C=0$. Taking these two steady state density profiles together, it is evident that chronic viruses should avoid latency ($f_C=0$), and therefore temperate viruses should adopt a lysogen frequency near $f_T=0.01$.} \label{fig:fitness}
\end{figure}

Figure \ref{fig:fitness}(a) shows the steady state temperate viral density for all possible combinations of $f_T$ and $f_C$. Given any chronic lysogen frequency $f_C$, temperate viruses may select a lysogen frequency $f_T$ that conditionally maximizes their viral density (optimal strategies are shown in red). We see that for all possible chronic latency strategies, there exists a small but nonzero (i.e., sweet spot) lysogen frequency $f_T$ that maximizes temperate viral density at steady state. 

\subsection{Chronic viruses should eliminate latency}
While lysogen frequencies for temperate viruses are well-studied, lysogen frequencies for chronic viruses are unknown. Our model reveals that the lysogen frequency for chronic viruses should be exactly zero. In Figure \ref{fig:fitness}(b), we plot the steady state density of chronic viruses in the system ($V_C+L_C$). For any given temperate viral lysogen frequency $f_T$, chronic viruses maximize steady state density by selecting a lysogen frequency $f_C=0$. From an ecological perspective, this result is intuitive. We have assumed that there is no reproductive cost to productive infection relative to latent infection, so it benefits chronic viruses to spread genetic material both horizontally (via production) and vertically (via cell division), rather than vertically alone.

\subsection{Bacterial recovery rate determinable using viral abundance}
We use our conclusion that $f_C=0$, along with the fact that temperate viruses possess a lysogen frequency around 1\% to deduce the typical bacterial recovery (cure) rates. In the interest of simplicity, we assume all recovery rates are equal: $\gamma=\gamma_T = \gamma_P = \gamma_C$. The appropriate recovery rates should lead to a maximum temperate virus steady state density ($V_T+L_T$) for $f_C = 0$ and $f_T \approx 0.01$, which occurs over only a small range of $\gamma \approx 0.67$; see Figure \ref{fig:fitness}(a).

\begin{figure}[h]
  \centering
\includegraphics[width=0.8\textwidth] {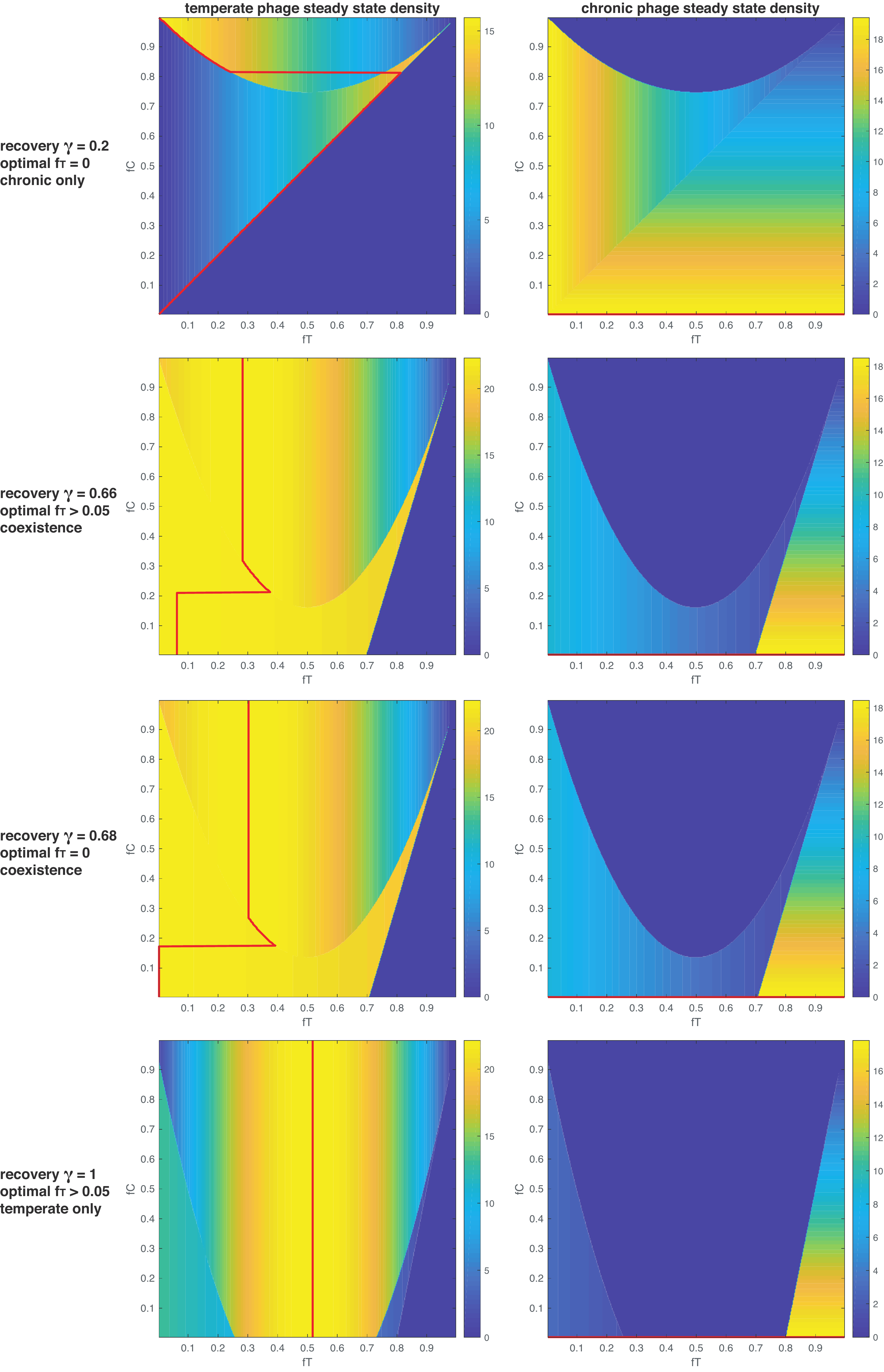}
      \caption{Viral steady state density over the full range of possible lysogen frequencies ($f_i\in[0,1]$) for temperate and chronic viruses. Color indicates the steady state viral density, and the red line is the maximum steady state density for each lysogen frequency of the competing virus. Left panel is temperate virus density ($V_T+L_T$) at steady state. Right panel is chronic virus density ($V_C+L_C$) at steady state. Top row is $\gamma=0.2$, followed by $\gamma=0.66, \gamma=0.68,$ and the bottom row is $\gamma=1$.} \label{fig:fitnessSI}
\end{figure}

A wide range of outcomes is possible if the recovery rate $\gamma$ is not near $0.67$ (see Figure \ref{fig:fitnessSI}), but none include temperate lysogen frequencies near 1\% and coexistence of both viral types, as we see in many natural environments. 

For $\gamma \le 0.2$ (very slow intracellular provirus deactivation), optimal temperate lysogen frequencies are $f_T=0$ when chronic viruses select the optimal lysogen frequency of $f_C=0$ (see Figure \ref{fig:fitnessSI}). However, temperate viruses are effectively driven to extinction under these conditions, implying that extremely stable proviruses are deleterious to temperate viruses. Due to the presence of both viral types in many environments, we suspect that intracellular provirus deactivation is not extremely slow.

For moderately slow intracellular provirus deactivation ($0.22\le \gamma \le 0.66$), optimal temperate lysogen frequencies exceed 5\% (see Figure \ref{fig:fitnessSI}). This result implies that temperate viruses more resilient to deactivation should also increase latency. 

If instead $0.68\le \gamma \le 0.79$ (moderately fast intracellular provirus deactivation), then the optimal temperate lysogen frequency is exactly zero again (see Figure \ref{fig:fitnessSI}). In other words, for faster recovery rates, all temperate viruses would be obligately lytic. Due to this result, we speculate that temperate proviruses are more resilient to intracellular deactivation than obligately lytic viruses. 

For $\gamma\ge 0.8$ (very fast intracellular deactivation), lysogen frequencies instantaneously jump to around $f_T=0.5$, and chronic viruses are driven to extinction (see Figure \ref{fig:fitnessSI}).

\section{Discussion}
\subsection{Recovery rates and lysogen frequencies}
In environments where temperate and chronic viruses coexist, bacterial recovery rates should be near $\gamma=0.67$. If proviruses are slightly more stable, then we would expect to see higher lysogen frequencies for temperate viruses. For proviruses that are stable on evolutionary timescales, we would expect to see chronic viruses dominate. For proviruses that are slightly less stable, we would expect to see the latency strategy disappear. For proviruses that are quickly deactivated, we would expect to see temperate viruses dominate with large lysogen frequencies. This predicted recovery rate is faster than one might expect, so we hope this study encourages more experimental work on the intracellular deactivation of proviruses within \textit{P.~aeruginosa}.

\subsection{Limitations}
With the goal of keeping the model analytically tractable, we have made simplifying assumptions that may affect the presented results. First, we have assumed mass action infection dynamics, but \textit{P.~aeruginosa}-virus infection rates may not be well-approximated by a mass action process, especially for large bacteria population sizes \cite{simmons2017phage,vidakovic2018dynamic}. More realistically, infection rates may slow as population growth activates quorum-sensing and biofilm formation \cite{harper2014bacteriophages}. 

In addition, we have assumed the lytic recovery rate is zero and all other recovery rates are equal. Although assuming $\gamma_I = 0$ produces qualitatively similar results to $\gamma_I = \gamma_T$, it may not be reasonable to assume that all other recovery rates are equal. Future study is needed to determine how recovery rates are affected by the infection type.

We have also assumed that both viral types produce super-infection and cross-infection exclusion proteins that prevent a second infection of any kind. While many viruses that infect \textit{P.~aeruginosa} produce super-infection exclusion proteins that effectively prevent multiple infections by the same viral type \cite{heo2007genome,james2012differential}, little is known about cross resistance to viral infection.

Another simplifying assumption is that lysogen frequencies are constant, but some viruses are able to detect bacteria population density, which appears to affect the frequency of lysogeny \cite{hargreaves2014does,silpe2018host}. If this process applies to \textit{P.~aeruginosa} and its viruses, a more sophisticated model would incorporate a density-dependent latency probability: $f_T(N)$ and $f_C(N)$. 

Finally, in deducing the expected chronic lysogen frequency $f_C$ and the recovery (cure) rates $\gamma$, we have assumed that all other parameters are exactly the baselines given in Table \ref{tab:param}. While the literature has provided reasonable ranges for these parameters, several baseline values (e.g., $\beta_C$, $\eta_T$, and $\eta_C$) were simply selected within those ranges. Given the uncertainty in several parameter values, the model-inferred parameters $f_C$ and $\gamma$ are also  uncertain.

\subsection{Future steps}
This model could serve as a base for more sophisticated extensions. For instance, the presented model does not include an evolutionary component and is therefore only applicable on short time scales. However, this model could be part of a multi-scale model that incorporates both short time-scale (ecological) dynamics and long time-scale (evolutionary) dynamics. 

Also, we have assumed that no environmental stressors (e.g., radiation, heat, sublethal antibiotics) perturb the system, but antibiotics are often used to treat bacterial infections. Many classes of antibiotics are known to induce latent proviruses and trigger virus production, even if the bacteria are antibiotic resistant \cite{rokney2008host,fothergill2011effect,lopez2014induction,martinez2015freeing,kaur2012plaque}. In fact, the induction of latent viruses is proposed to be one of the mechanisms behind the synergistic effect of antibiotics and viruses to treat recalcitrant bacterial infections \cite{kaur2012plaque,kim2018phage}. Infections by \textit{P.~aeruginosa} represent about 10\% of nosocomial infections, are a leading cause of death among patients with cystic fibrosis, and have been deemed a serious threat on the United States Centers for Disease Control watch list for antibiotic resistance \cite{jarvis1992predominant,hancock2000antibiotic,us2013antibiotic}; therefore, a critical next step is understanding the impact of antibiotic-induced proviruses on control of bacterial infections. This is the subject of ongoing study.

\section{Conclusion}
We have developed a simple mathematical model of the ecological competition between temperate and chronic viruses for bacterial hosts. Using the hosts \textit{E.~coli} and \textit{P.~aeruginosa} as motivation, we demonstrate that low lysogen frequencies provide competitive advantages for both viral types. Interestingly, chronic viruses theoretically maximize their steady state density by eliminating latency entirely, but temperate viruses exhibit a non-zero `sweet spot' lysogen frequency. Using experimental evidence that temperate viruses possess lysogen frequencies around 1\% and that both viral types coexist in real environments, we are able to estimate the recovery (cure) rates for bacteria. Better understanding of this system may contribute to optimal treatment of bacterial infections using phage therapy and/or antibiotics.

\section{Acknowledgements}
This work was funded in part by the National Science Foundation grant DMS-1815764 (ZR), the Cystic Fibrosis Foundation grant WHITAK16PO (RJW), and an Allen Distinguished Investigator Award (RJW). The funders had no role in study design, data collection and analysis, decision to publish, or preparation of the manuscript.

The authors thank Jayadevi H.~Chandrashekhar, Ted Kim, and George A.~O'Toole for discussions that informed biological aspects of this work. 

\section{Data availability}  
All software (Matlab .m files) will be made publicly available via the Illinois Data Bank pending publication in a peer-reviewed journal.

\section{Competing interests}
The authors declare no competing interests.

\bibliographystyle{ieeetr}
\bibliography{pseudomonasLibrary_v4}

%


\setcounter{section}{0}
\renewcommand{\thesection}{S\arabic{section}}

\newpage

\footnotesize
\section{Supplementary Information}
\subsection{Steady states}
The following biologically relevant steady-states exist for system 
(\ref{eq:sus}-\ref{eq:chronvir}).
\begin{itemize}

\item
First, it can be readily seen that the trivial steady state, where all population densities are zero, exists.

\item
Second, the steady state where only the susceptible class persists and is equal to 
the bacterial carrying capacity, $S = K$, also exists for all parameter values.
 
\item
The steady state where only the temperate phage persists, namely 
$P_C = V_C = L_C = 0$, also exists. In this case it holds:
\begin{align*}
& S= \frac{\mu_T}{\eta_T} \frac{\delta+\gamma_I}
{\beta_T \delta (1-f_T) -(\delta+\gamma_I)}\\
& V_T = \frac{\beta_T \delta(1-f_T) -(\delta+\gamma_I)}{\mu_T (1-f_T)} I_T \\
& I_T = \frac{r_S S (K-S)(1-f_T)}{(r_S S -\gamma_I K)(1-f_T) + (\delta+\gamma_I) K} +
\frac{(1-f_T) (\gamma_T K-r_S  S)}{(r_S S-\gamma_I K) (1-f_T) + (\delta+\gamma_I) K} L_T 
\\
& r_T \left((\gamma_T-\gamma_I )(1-f_T) + \delta + \gamma_I \right) L_T^2 + 
(\gamma_T (1-f_T)(r_S S +\delta K) + (\delta+\gamma_I) r_S f_T S - (\delta+f_T \gamma_I) r_T (K-S)) L_T -\\
& f_T (\delta+\gamma_I) r_S S (K-S) =0
\end{align*}

\item
Similarly, when only the chronic phage persists, namely when $I_T = V_T = L_T = 0$, it holds
\begin{align*}
& S=  \frac{\mu_C}{\eta_C} \frac{\gamma_P}{\beta_C(1-f_C) - \gamma_P} \\
& P_C = \frac{K-S}{1+\frac{f_C \gamma_P}{(1-f_C) \gamma_C}} \\
& L_C = \frac{f_C \gamma_P}{(1-f_C) \gamma_C} P_C \\
& V_C = \frac{\beta_C (1-f_C)-\gamma_P}{\mu_C (1-f_C)} P_C
\end{align*}

\item
Finally, if all population densities are positive, it holds
\begin{align*}
& S= \frac{\mu_T}{\eta_T} \frac{\delta+\gamma_I}{\beta_T \delta (1-f_T) -(\delta+\gamma_I)}\\ 
& 1-\frac{N}{K} = \frac{\gamma_P}{r_P} - \frac{(1-f_C)\beta_C}{r_P} \frac{\eta_C S}{\eta_C S + \mu_C} \\
& V_T = \frac{\beta_T \delta(1-f_T) -(\delta+\gamma_I)}{\mu_T (1-f_T)} I_T \\
& V_C = \frac{\beta_C}{\eta_CS + \mu_C} P_C \\
& L_T = \frac{f_T}{1-f_T} \frac{\delta + \gamma_I}{\gamma_T - r_T \left( 1-\frac{N}{K} \right)} I_T \\
& L_C = \frac{\eta_C f_C \beta_C S}{\eta_C S+\mu_C} 
\frac{1}{\gamma_C - r_C \left( 1-\frac{N}{K} \right)} P_C. 
\end{align*}
The following linear $2 \times 2$ system can be solved to yield a unique steady-state.
\begin{align*}
&  \left( \frac{\delta+\gamma_I}{1-f_T} 
\frac{\gamma_T (1-f_T) -r_T \left( 1-\frac{N}{K} \right)}{\gamma_T -r_T \left( 1-\frac{N}{K} \right)} -\gamma_I \right) I_T + 
\left( \frac{\eta_C \beta_C S}{\eta_C S+\mu_C} 
\frac{\gamma_C (1-f_C) -r_C \left( 1-\frac{N}{K} \right)}{\gamma_C -r_C \left( 1-\frac{N}{K} \right)} -\gamma_P \right) P_C
  = \\
& r_S S \left( 1-\frac{N}{K} \right)\\ 
& \left(1+  \frac{f_T}{1-f_T} \frac{\delta + \gamma_I}{\gamma_T - r_T \left( 1-\frac{N}{K} \right)} \right) I_T + 
\left( 1+ \frac{\eta_C f_C \beta_C S}{\eta_C S+\mu_C} 
\frac{1}{\gamma_C - r_C \left( 1-\frac{N}{K} \right)} \right) P_C = N-S
\end{align*}

\end{itemize}

\subsection{Viral invasion fitness}
The basic reproductive number $R_0$, traditionally defined as the average number of new infections 
generated by an infectious individual in an entirely susceptible population
\cite{diekmann1990}, has long been used to characterize a pathogen's fitness 
\cite{rapti2016,weitz2019viral,andreasen1995pathogen,adams2007cross}. The basic reproductive number has been found to be correlated with the
between-host transmission rate \cite{beretta1998}, the number of 
free infective propagules produced per infected host (virions, in our case), and life-history traits such as mortality, fecundity, and growth \cite{rapti2016,gilchrist2006}.

In this study, $R_0$ is used to determine the growth rate of a viral invader in a population of residents at steady state. Specifically, when either the temperate or chronic virus attempts to invade the 
bacterial population, the virus is successful when its respective $R_0$ is greater 
than one. Similarly, when one of the two viruses is the resident, the other virus can invade and 
coexist as long as its $R_0$ is greater than one. The various regions where 
each virus invades and persists are shown in Figure \ref{fig:bif3}.  

One of the goals and challenges of phage therapy is to ensure that there is active viral replication
\cite{payne2001}. This occurs when, after an initial dose of the virus, it is able to proliferate in its 
bacterial host. There are cases however, when viral replication does not 
take place, so repeated administration of the virus is required. There already exist various models 
that study viral kinetics and provide thresholds that guarantee active replication of lytic-only viruses 
\cite{payne2001,payne2003}. Yet, as in most models, recovery of the bacterial host is neglected.

Given the large range of recovery rates that viruses exhibit \cite{brussow2004phages,horvath2010crispr}, 
in this work, we investigate and identify viral recovery strategies that optimize viral abundance. Although the basic reproductive number is a traditional measure of fitness (see e.g., \cite{wahl2019evolutionary}), using abundance as a proxy for competitive advantage is also common (see e.g., \cite{thingstad2014theoretical}). 

\subsection{Linear stability analysis and bifurcations}

\begin{enumerate}
\item The trivial equilibrium is linearly unstable for all choices of 
parameter values.
\item The steady-state with $S=K$ is linearly stable as long as
\[R_T =\frac{(1-f_T) \eta_T \beta_T \delta K}{(\mu_T + \eta_T K)(\delta+\gamma_I)} <1,~~~
R_C = \frac{(1-f_C) \eta_C \beta_C K}{\gamma_P (\mu_C + \eta_C K)} <1.\] 
We notice from the previous section that when $S=K$ it holds
\[ S =\frac{\mu_T}{\eta_T} \frac{\delta+\gamma_I}
{\beta_T \delta (1-f_T) -(\delta+\gamma_I)} = K \Leftrightarrow 
R_T =1\]
and
\[ S=  \frac{\mu_C}{\eta_C} \frac{\gamma_P}{\beta_C(1-f_C) - \gamma_P} =K \Leftrightarrow
R_C=1.\]
Therefore, the bacterium only ($S=K$) steady-state undergoes a transcritical 
bifurcation with the $P_C=V_C=L_C=0$ (temperate only) steady state when $R_T=1$. 
Similarly, it undergoes a transcritical bifurcation with the $I_T=V_T=L_T=0$ (chronic only) 
steady state when $R_C=1$.
\item The temperate only steady state undergoes a transcritical bifurcation with the 
coexistence steady state when  
\begin{align*}
R_{TC} = \frac{(1-f_C) \beta_C \eta_C S}{\eta_C S + \mu_C} 
\frac{1}{\gamma_P - r_P \left(1 -\frac{N}{K} \right)}=1, ~~~\mbox{where}~~~
S= \frac{\mu_T}{\eta_T} \frac{\delta+\gamma_I}
{\beta_T \delta (1-f_T) -(\delta+\gamma_I)},
\end{align*}
and $N=S+I_T+L_T$.
\item There is a transcritical bifurcation from the chronic only to the 
coexistence steady-state when     
\begin{align*}
R_{CT} = \frac{(1-f_T) \beta_T  \delta \eta_T S}{(\eta_T  S + \mu_T)(\delta+\gamma_I)}=1, ~~~
\mbox{where}~~~ S = \frac{\mu_C}{\eta_C} \frac{\gamma_P}{\beta_C(1-f_C) - \gamma_P}.
\end{align*}
\end{enumerate}
All these bifurcations are obtained by a standard linear stability analysis around 
the relevant steady states.

\subsection{Model non-dimensionalization}
The model system (\ref{eq:sus}-\ref{eq:chronvir}) can be non-dimensionalized so that time and bacterial density are unitless quantities. We will illustrate with equation (\ref{eq:sus}) with $\gamma_I=0$, but the methodology is analogous for all other equations in the system. Suppose the system (\ref{eq:sus}-\ref{eq:chronvir}) has bacterial density units of CFU/mL and time units of minutes. Then equation (\ref{eq:sus}) has units CFU/mL/min:

\begin{equation*}
\frac{\mathrm{d} S}{\mathrm{d} t} =  \underbrace{r_S S \left( 1-\frac{N}{K} \right)}_{\text{growth}} - \underbrace{\eta_T S V_T - \eta_C S V_C}_{\text{infection}} + \underbrace{\gamma_T L_T+ \gamma_P P_C + \gamma_C L_C}_{\text{recovery}}.
\end{equation*}

We will divide the entire equation by $r_S K$, i.e.~the bacterial growth rate (in min$^{-1}$) multiplied by the carrying capacity (in CFU/mL):

\begin{equation*}
\frac{\mathrm{d} (S/K)}{\mathrm{d} (r_S t)} =  \underbrace{\frac{S}{K} \left( 1-\frac{N}{K} \right)}_{\text{growth}} - \underbrace{\frac{\eta_T}{r_S} \frac{S}{K} V_T - \frac{\eta_C}{r_S} \frac{S}{K} V_C}_{\text{infection}} + \underbrace{\frac{\gamma_T}{r_S} \frac{L_T}{K}+ \frac{\gamma_P}{r_S} \frac{P_C}{K} + \frac{\gamma_C}{r_S} \frac{L_C}{K}}_{\text{recovery}}.
\end{equation*}

Now $\tilde{S}=S/K$ is a unitless bacterial density, $\tilde{t}=r_S t$ is a unitless time, $\tilde{\eta_i}=\eta_i/r_S$ is an infection rate per PFU/mL, and $\tilde{\gamma_i}=\gamma_i/r_S$ is a unitless recovery rate: 
\begin{equation*}
\frac{\mathrm{d} \tilde{S}}{\mathrm{d} \tilde{t}} =  \underbrace{\tilde{S} \left( 1-\tilde{N} \right)}_{\text{growth}} - \underbrace{\tilde{\eta}_T \tilde{S} V_T - \tilde{\eta}_C \tilde{S} V_C}_{\text{infection}} + \underbrace{\tilde{\gamma}_T \tilde{L}_T+ \tilde{\gamma}_P \tilde{P}_C + \tilde{\gamma}_C \tilde{L}_C}_{\text{recovery}}.
\end{equation*}

For the sake of clarity, we have suppressed tilde notation throughout the main manuscript. The same effect is achieved by choosing $\gamma=1$ and $K=1$ in system (\ref{eq:sus}-\ref{eq:chronvir}) and reinterpreting bacterial populations as fractions of the carrying capacity and rates (except infection and adsorption) as multiples of the growth rate. All parameters in Table II have been scaled as such (details follow).

\subsection{Parameter selection}
The \textbf{growth rate} $r_S$ for \textit{P.~aeruginosa in vitro} is approximately 5.1e-3 min$^{-1}$ \cite{spalding2018mathematical}, although \textit{P.~aeruginosa} growth is highly variable in humans \cite{kopf2016trace}. Therefore all rate parameters provided in min$^{-1}$ are scaled by this rate in order to non-dimensionalize.

The \textbf{carrying capacity} $K$ of bacteria in a medium depends on the environment. Even within the sputum of patients with cystic fibrosis, the carrying capacity is difficult to estimate due to variability among patients. One study of patients with cystic fibrosis found that the densities of viable \textit{P.~aeruginosa} in sputum of 12 patients not undergoing treatment ranged from 5.3e3 CFU/mL to 1.8e11 CFU/mL \cite{stressmann2011does}. 

The \textbf{infection rate} for \textit{E.~coli} and T4 phage in mucus (assuming mass action infection) is known to be approximately 47e-10 mL/min per CFU per PFU \cite{stent1963molecular,barr2015subdiffusive}. Infection in marine ecosystems (also assuming mass action infection) is similar at around 24e-10 mL/min per CFU per PFU \cite{stent1963molecular,thingstad2014theoretical}. Given our uncertainty in the bacterial carrying capacity, we elected to use an infection rate within the range given by Sinha et al. \cite{sinha2017silico}; the authors fit their mass action infection model to time series population data that reached carrying capacity. The authors presented $K\eta \in [0.45, 100]$ hr$^{-1}$ and $r_S\in[0.5, 10]$ hr$^{-1}$. Non-dimensionalization leads to a range of $\eta$ between $0.045$ and $200$, and our selected value of $\eta=1$ is near the geometric mean of that range. 

The \textbf{phage production delay rate} $\delta$ is estimated based on the eclipse and rise phase of PAXYB1 and PAK\_P3 phage \cite{yu2017characterization,el2015isolation}. The eclipse (latent) and rise phase is 130 minutes total for PAXYB1 \cite{yu2017characterization} and 27 minutes total for PAK\_P3 \cite{el2015isolation}. The smallest (non-dimensional) delay rate is then $1/130/$5.1e-3$=1.5$, and the largest is $1/27/$5.1e-3$=7.3$. We selected the approximate average of this range, 4, to be the delay rate.

The \textbf{phage degradation rate} $\mu_i$ is estimated based on the decay rates of phage in aquatic environments \cite{heldal1991production}. The decay rates ranged from 0.26 to 1.1 per hour. We non-dimensionalize by multiplying by 60 minutes per hour and the bacterial growth rate, 5.1e-3 per minute. The non-dimensional range of decay rates is then 0.9 to 3.6. We selected a value of 1 arbitrarily from this range.

\end{document}